**Melting point depression of charge density wave in 1*T*-TiSe₂ due to size effects**

Saif Siddique[1,2,†], Mehrdad T. Kiani[1,3,†], Omri Lesser[4], Stephen D. Funni[1], Nishkarsh Agarwal[5], Maya Gates[5,6], Miti Shah[5], William Millsaps[5], Suk Hyun Sung[7], Noah I. Schnitzer[1,2], Lopa Bhatt[8], David A. Muller[2,8], Robert Hovden[5], Ismail El Baggari[7], Eun-Ah Kim[4], Judy J. Cha[1,*]

[1] Department of Materials Science and Engineering, Cornell University, Ithaca, NY 14853, USA
[2] Kavli Institute at Cornell for Nanoscale Science, Cornell University, Ithaca, NY 14853, USA
[3] Department of Materials Science and Engineering, FAMU-FSU College of Engineering, Tallahassee, FL 32310, USA
[4] Department of Physics, Cornell University, Ithaca, NY 14853, USA
[5] Department of Materials Science and Engineering, University of Michigan, Ann Arbor, MI 48109, USA
[6] h-Bar Instruments, Ann Arbor, MI 48103, USA
[7] The Rowland Institute at Harvard, Harvard University, Cambridge, MA 02138, USA
[8] School of Applied and Engineering Physics, Cornell University, Ithaca, NY 14853, USA

[†] Equal Contribution
[*] Corresponding author: jc476@cornell.edu

**Abstract**

Classical nucleation theory predicts size-dependent nucleation and melting due to surface and confinement effects at the nanoscale. In correlated electronic states, observation of size-dependent nucleation and melting is rarely reported, likely due to the extremely small length scales necessary to observe such effects for electronic states. Here, using 1*T*-TiSe₂ nanoflakes as a prototypical two-dimensional (2D) charge density wave (CDW) system, we perform *in-situ* cryogenic electron microscopy with temperature down to 20 K and observe size-dependent nucleation and melting of CDWs. Specifically, we observe a melting point depression of CDW for 1*T*-TiSe₂ flakes with lateral sizes less than 100 nm. By fitting experimental data to a Ginzburg-Landau model, we estimate a zero-temperature correlation length of 10–50 nm, which matches the reported CDW domain size for 1*T*-TiSe₂. As the flake size approaches the correlation length, the divergence of the CDW correlation length near the transition is cut off by the finite flake size, limiting long-range order and thereby lowering the transition temperature. For very small flakes whose size is close to the correlation length, we also observe absence of CDWs, as predicted by the model. We thus show that an electronic phase transition follows classical nucleation theory.

**Introduction**

Electronic states in nanoscale samples can deviate from bulk behaviors due to surface and confinement effects when the sample size approaches the governing temperature-dependent correlation length, $\xi(T)$, of the electronic states. For example, superconducting nanowires with diameters approaching the size of Cooper pairs exhibit a decrease in their superconducting transition temperature with complete suppression of superconductivity below $\xi(T)$ due to quantum phase slips.[1,2] Similar size-dependent phase instability is observed in a wide variety of electronic states including excitonic insulators[3] and quantum Hall edges.[4] Thus, $\xi(T)$ sets a fundamental size limit for a given quantum state, which is critical for microelectronic and quantum device applications that seek to incorporate quantum materials.

In charge density wave (CDW) systems, two length scales are relevant: the correlation length over which CDW fluctuations are correlated and the domain size.[5] The correlation length depends on the temperature: finite at zero temperature and divergent near the transition temperature. The zero-temperature correlation length $\xi_0 \sim \frac{h v_F}{\pi \Delta}$, where $h$ is the Planck's constant, $v_F$ is the Fermi velocity and $\Delta$ is half the band gap opening due to the CDW formation,[5] reflects the balance between quantum fluctuations due to the kinetic energy and the interaction driving the CDW ordering. The zero-temperature correlation length $\xi_0$ in weak coupling systems can approach several microns, and it is much larger than the lattice constant $a$. In contrast to 1D CDW systems, most 2D CDW systems exhibit stronger electron-phonon coupling and a larger band gap, leading to a much shorter zero-temperature correlation length $\xi_0$ of 1–20 nm.[6] For layered CDW compounds, such as 2T-TaSe$_2$, 1T-TaS$_2$, 2H-NbSe$_2$, and 1T-TiSe$_2$, precise knowledge of the zero-temperature correlation lengths and nanoscale CDW stability is important as they have shown promise for applications in switchable electronics due to their metal-insulator transitions and I-V hysteresis.[7–9]

The second length scale, the CDW domain size, corresponds to regions of the sample that have the same periodic electronic and structural distortions. Unlike the intrinsic property of a clean system $\xi_0$, the domain size can depend on extrinsic factors, such as synthesis and processing conditions. However, the domain size at a given temperature $T$ cannot be smaller than $\xi(T)$. Studies have shown that CDW domain size corresponds to spacing between structural disorder such as crystalline defects or intercalants in 2D materials.[10–13] These defects act as preferential sites for CDW nucleation, forming domain walls between adjacent CDW domains. Since the domain size can depend on extrinsic factors, there is a broad size distribution for CDW domains. In 1T-TaS$_2$, CDWs nucleate in concentrated regions of basal dislocations and domain size can range from 5 nm up to many microns.[14,15] In 1T-TiSe$_2$, domains are centered about self-intercalated Ti clusters and the spacing between clusters can range between several nanometers up to 60 nm.[16]

For nanoflakes whose lateral dimensions approach either the domain size or $\xi(T)$, confinement and edge effects are expected to suppress CDW formation and stability. To date, experimental studies have focused on thickness dependence using laterally large exfoliated flakes,[17,18] larger than typical CDW domain sizes,[19] or low-temperature correlation length ~$\xi_0$[20] and importantly, larger than the size of modern devices. Studies on samples with lateral dimensions approaching relevant length scales have been inconclusive regarding CDW behavior. In TaS$_3$, a quasi-1D van der Waals (vdW) material with a unidirectional CDW with $\xi_0$ = 10 μm, very weak CDW melting point suppression was observed in wires shorter than $\xi_0$.[21] For 1T-NbSe$_2$ with $\xi_0$ = 20 nm,[20] scanning tunneling microscopy (STM) at 4.2 K shows stable CDW order in 10 nm diameter 1T/1T bilayer NbSe$_2$ flakes, but CDW melting in 1$T$/2$H$ bilayer flakes below 26 nm in diameter.[22] For 2D transition metal dichalcogenides that are prone to the formation of chalcogen vacancies and self-interstitials during growth,[16,23] stability of their CDW order may get modified significantly.

Here, we experimentally observe lateral confinement effects on CDW melting at sub-micron length scales using *in-situ* liquid helium and liquid nitrogen electron microscopy. We study nanoflakes of 1$T$-TiSe$_2$, a prototypical CDW material with CDW transition temperature ranging between 210 K and 230 K for bulk samples.[17] Previous studies on 1$T$-TiSe$_2$ report domain size of 10–35 nm in-plane and 17–30 nm out-of-plane.[19] We observe a significant CDW melting point depression in nanoflakes of effective radii < 100 nm, and a complete suppression of CDW transition in nanoflakes of sizes ~50 nm. Fitting the observed melting point depression using a Ginzburg-Landau model results in a zero-temperature correlation length of 10–50 nm. Finally, using atomic-resolution imaging we show that Ti self-intercalant clusters, which act as CDW nucleation sites,[16] are spaced within this same length scale. Thus, our observation demonstrates size-dependent melting point depression and instability for CDW in 1$T$-TiSe$_2$.

**Results**

1$T$-TiSe$_2$ is a layered transition metal dichalcogenide that belongs to the $P\bar{3}m1$ space group at room temperature. Upon cooling, bulk 1$T$-TiSe$_2$ undergoes an electronic phase transition into a commensurate 2×2×2 CDW phase below 210–230 K. This CDW transition is accompanied by a periodic lattice distortion where titanium (Ti) and selenium (Se) atoms are displaced from their equilibrium sites, modulating the Ti-Se bonds and doubling the lattice periodicity.[19] The driving mechanism behind the CDW transition is believed to be excitonic condensation, where electron-hole pairs stabilize the CDW state.[24,25]

While some CDW systems are sensitive to sample thickness—for example, suppression of the commensurate CDW phase in 1$T$-TaS$_2$ with thicknesses < 20 nm[26] and change of the CDW transition temperature in thin flakes of GdTe$_3$[27]—the CDW transition in 1$T$-TiSe$_2$ is robust against variations in sample thickness, persisting down to monolayer without large deviations in the transition temperature.[24] This makes 1$T$-TiSe$_2$ a suitable system to study the effects of lateral confinement on the stability of CDW, which has not been systematically explored in 2D CDW systems.

To study the lateral size effects on the CDW of 1$T$-TiSe$_2$, we synthesize nanoflakes with varying sizes, focusing on flakes with effective radii < 250 nm. We first synthesize TiSe$_2$ bulk powder following a solid-state synthesis approach.[28] Powder X-ray diffraction confirms the final product to be TiSe$_2$ with residual Ti$_3$Se$_4$ (< 5%) (Supplementary Figure 1). Nanoscale flakes were then obtained by liquid exfoliation of the powder dispersed in methanol via ultrasonication (synthesis details in Supplementary Note 1). The dispersion was dropcast onto TEM grids for *in-situ* cryogenic electron microscopy. Figure 1a shows a bright-field (BF) TEM image of 1$T$-TiSe$_2$ flakes of different sizes. We examined nanoflakes with projected surface areas < 200,000 nm$^2$ (effective radii < 250 nm, assuming circular flakes) to study the size effects.

Atomic-resolution high-angle annular dark-field scanning transmission electron microscopy (HAADF-STEM) images of the TiSe$_2$ nanoflakes confirm the atomic structure of the 1$T$ phase (Figure 1c), matching that of a commercially purchased 1$T$-TiSe$_2$ single crystal (Supplementary Figure 2).

As TiSe$_2$ is known to be slightly air sensitive,[29] we test the oxidation state of the TiSe$_2$ nanoflakes using HAADF-STEM (Figure 1b) and electron energy loss spectroscopy (EELS) (Figures 1d–f). The interior of the flake shows distinct atomic columns reflecting the 1$T$ phase while the edge of the flake shows a 2–4 nm wide amorphous region. We acquired EELS mapping with Ti-L$_{2,3}$ and O-K edges (Fig. 1d) to generate Ti and O elemental maps (Figures 1e,f), which show a higher oxygen content at the flake edge compared to its interior. Thus, we conclude that the thin amorphous region at the flake edge is amorphous titanium oxide (TiO$_x$). We did not observe any flake size dependence on the oxide thickness or width, which remained 2–5 nm wide irrespective of the flake size (Supplementary Figure 3).

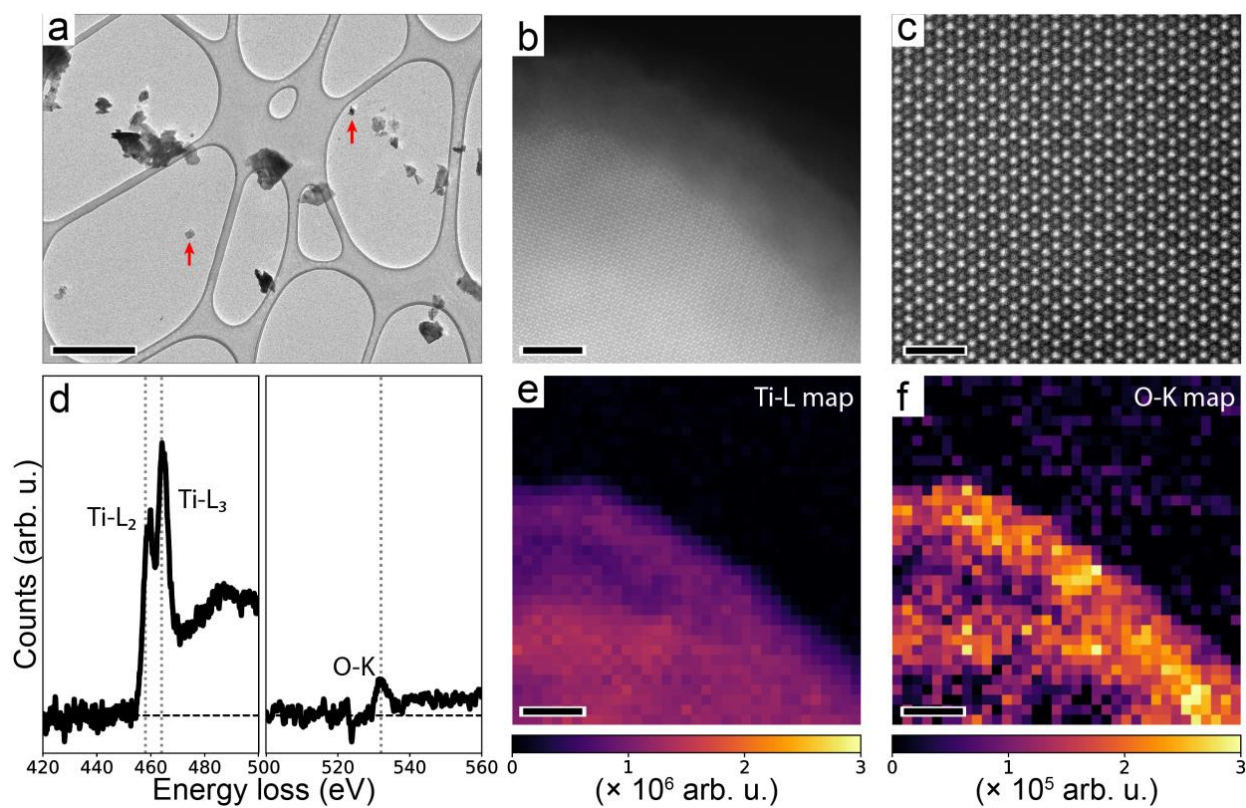

**Figure 1: Structure and stability of 1$T$-TiSe$_2$ nanoflakes**. **a.** BF-TEM image of TiSe$_2$ flakes of different sizes. Red arrows indicate example flakes with area below 200,000 nm$^2$. **b.** HAADF-STEM image of flake edge showing thin amorphous layer with **(c)** atomic resolution image from a different region of the nanoflake in **(b)**. **d.** EEL spectra from the flake, with intensity color maps

of **(e)** Ti-L peak and **(f)** O-K peak. Note we increased the intensity of the O-K map by a factor of 10 for visual clarity. Scale bars for **(a)** = 500 nm, for **(b, e, f)** = 3 nm, and for **(c)** = 1 nm.

Next, we examine the stability and evolution of the CDW in TiSe$_2$ as a function of lateral size by tracking TiSe$_2$ nanoflakes, with effective radii < 250 nm. We first cool the flakes to base temperature (either 20 K or 100 K using liquid helium or liquid nitrogen cryo-TEM holders, respectively) to induce the CDW phase. The ultra cold liquid helium cryo-TEM holder uses continuous liquid helium flow to provide high stability across intermediate temperatures by regulating flow rate and temperature at the heat exchanger.[30] The temperature was then gradually increased in steps of 10–20 K to room temperature. During the temperature increase, nanobeam electron diffraction patterns are collected in the <001> zone-axis of TiSe$_2$ using a ~ 5 nm-wide electron probe. At each temperature, 2x2 superlattice peaks are analyzed to track the melting of CDW, which is determined by the disappearance of the superlattice peaks above a certain temperature. The TEM is operated at an accelerating voltage of 60 kV to minimize sample damage due to electron irradiation that can induce a periodic distortion in the Ti atomic positions and result in superlattice peaks at the same position as CDW peaks in electron diffraction (Supplementary Figure 4).[31] If the TiSe$_2$ flakes undergo the irreversible irradiation damage, the 2x2 superlattice peaks persist even at room temperature. As we cannot distinguish between room temperature CDW fluctuations[32] and the irradiation damage, we exclude flakes that show superlattice peaks at room temperature from our analysis.

Figure 2 shows a BF-TEM image of a nanoflake with a projected surface area of ~ 6200 nm$^2$, with no observable structural defects. From the base temperature of 20 K to room temperature, we acquire nanobeam electron diffraction patterns from this flake. The CDW in this nanoflake melts between 70 K to 90 K, as indicated by the disappearance of the superlattice peaks (black arrows in Fig. 2). This temperature is well below the CDW melting temperature of ~ 210 K for bulk TiSe$_2$ (Supplementary Figure 5). Supplementary Figure 6 shows results of two additional nanoflakes, studied using the liquid nitrogen TEM holder. The melting temperatures of the CDWs for these two flakes are 170 K and 205 K having surface areas of ~ 3100 and ~ 19000 nm$^2$, respectively, again demonstrating melting point depression due to lateral confinement. We note that the intensity changes in Bragg peaks in Fig. 2 are due to sample motions, such as bending and tilting, due to the thermal drifts of the stage during temperature increase.

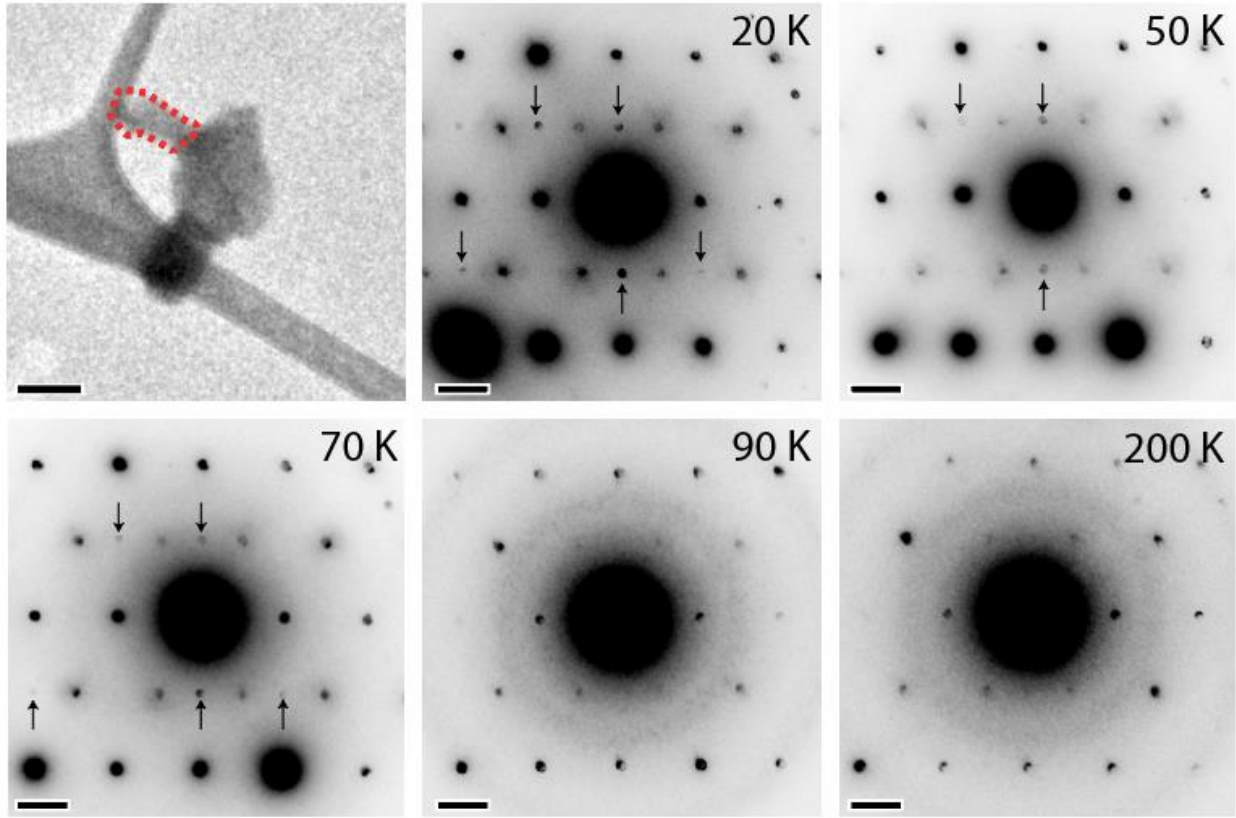

**Figure 2: CDW melting in 1$T$-TiSe$_2$ nanoflake by electron diffraction.** BF-TEM image of a nanoflake (dotted outline) with corresponding nanobeam electron diffraction patterns at different temperatures. Black arrows highlight prominent CDW peaks that disappear at higher temperatures. The overlapping nanoflakes in the BF-TEM image were separated by assuming each nanoflake has a convex shape. The concave points in the image were identified as junctions where these nanoflakes overlap. Scale bars for BF-TEM image = 100 nm, for diffraction patterns = 2 nm$^{-1}$.

Figure 3 summarizes the size-dependent CDW melting temperature ($T_{melt}$) of TiSe$_2$ nanoflakes. To quantify the CDW melting and get an accurate value for $T_{melt}$, we compare the CDW peak intensity to the background diffraction signal. When the peak intensity to background ratio falls below 5, the CDW is considered to have melted. A systematic decrease in $T_{melt}$ is clearly observed, starting from the effective radius of 100 nm (or projected surface area < 30,000 nm$^2$). The decrease in $T_{melt}$ becomes more significant as the flake becomes smaller. For nanoflakes with effective radii < 60 nm, $T_{melt}$ decreases by over 50 K compared to bulk. Because the diffraction patterns were taken at 10–20 K temperature steps, we represent the CDW melting range using lower- and upper-bounds corresponding to the temperature steps. Nanoflakes with effective radii larger than 100 nm show no change in $T_{melt}$ from bulk.

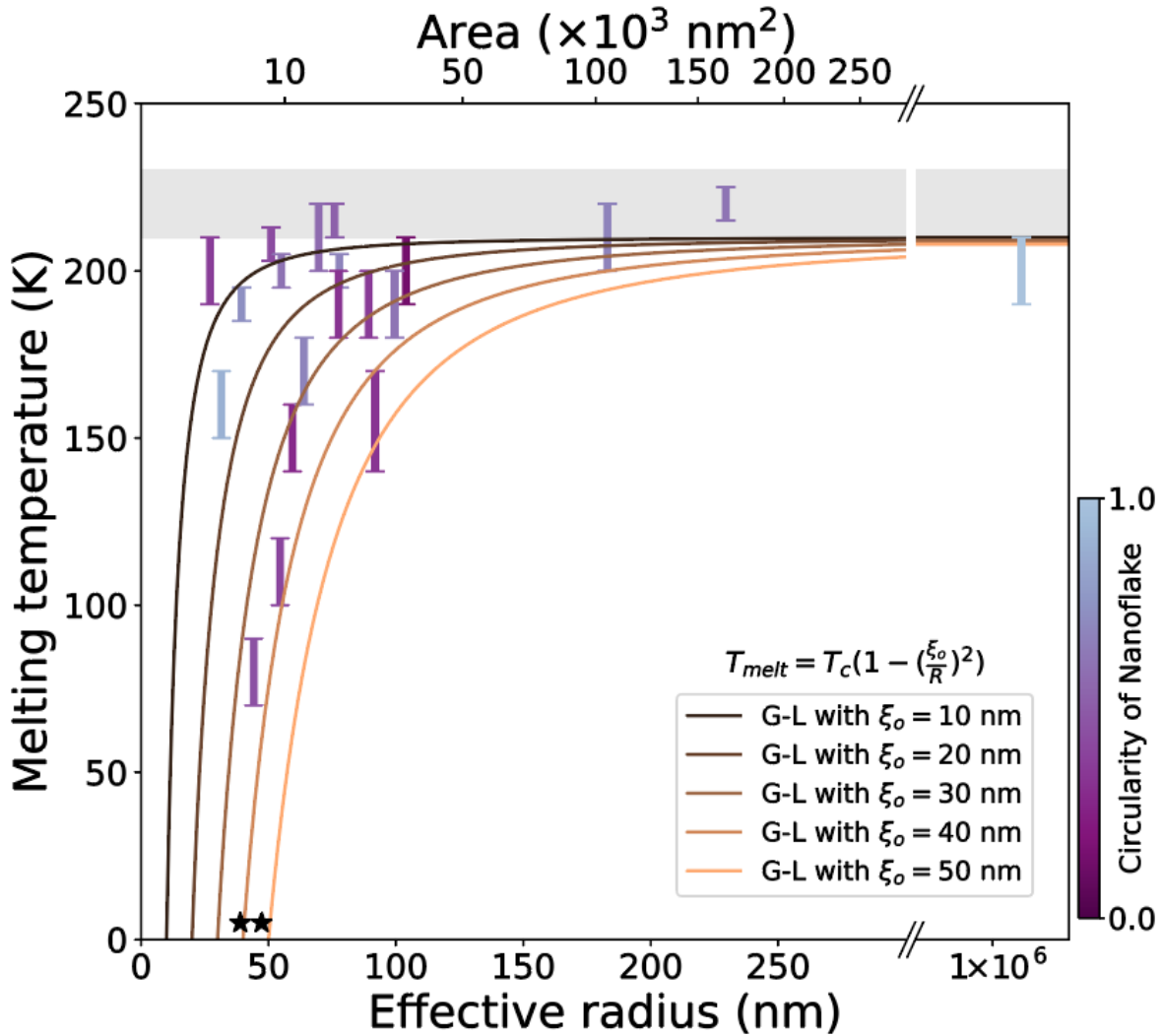

**Figure 3: CDW melting temperature vs size of nanoflakes.** Data points are represented by short vertical segments that reflect uncertainty with temperature measurement due to temperature increments of 10–20 K. The bounds show the temperature range between which the CDW melts. The upper x-axis shows the measured projected surface area of the flakes, and the lower x-axis shows the calculated effective radius of the flakes, assuming a circular flake of the same projected surface area. To account for the shape factor of nanoflakes, the data points are color-coded with circularity—value of 1 representing a circular flake. Overlaid on the data are the fitted results from the G-L calculations, which show a similar trend of suppression in the CDW transition temperature for finite-size system. Shaded region denotes the reported bulk transition temperature of 210–230 K.[17] The stars represent two flakes that did not transition into the CDW phase even at 20 K.

The onset of the CDW melting point depression in TiSe$_2$ nanoflakes occurs at an effective radius of ~ 100 nm, which is larger than the reported values for CDW domain size and correlation length in TiSe$_2$.[16,19] To understand the observed behavior, we model the CDW melting in a finite-size system of radius $R$ using Ginzburg-Landau (G-L) theory (see Supplementary Note 2 for calculation details). We consider a scalar order parameter $\psi$ for the commensurate CDW, with a free energy that contains a quadratic term, a quartic term, and a gradient term. The gradient term enables us to take finite size effects into account: by enforcing vanishing of the order parameter at the boundary and requiring that it be continuous, we model its decay from the bulk. This phenomenological theory describes the phase transition as a function of the system size normalized by its internal length scale, which is the correlation length. After minimizing the free energy functional, we get a relation between the radius of the system $R$ and the CDW melting temperature of the system $T_{melt}(R)$:

$$T_{melt}(R) = T_c\left(1 - \left(\frac{\xi_0}{R}\right)^2\right)$$

where $T_c$ is the CDW transition temperature at $R \to \infty$ and $\xi_0$ is the zero-temperature correlation length. With $T_c = 210\ K$ for 1$T$-TiSe$_2$, the modelled $T_{melt}(R)$ decreases with decreasing $R$ (Supplementary Figure 7), in agreement with the experimental results.

Figure 3 shows that the experimentally measured CDW melting temperatures have a considerable spread as a function of size. The size of the nanoflakes is obtained by measuring the area of the nanoflakes from TEM images and then using the area to get the equivalent circular radius. This approach however ignores the shape of the flake. The large spread in $T_{melt}$ values may partly be attributed to this shape factor of the nanoflakes. To account for this, we define a circularity metric, which is the ratio of smallest to the largest distances between two parallel tangents to the nanoflakes' boundaries (i.e., Feret's diameter). If this ratio is close to 1, the flake is more circular, but if the ratio is close to 0, the flake shape is highly elongated. We color-code the data points in Fig. 3 (Supplementary Figure 8 shows the minimum and maximum Feret's radii on the $T_{melt}$ vs size plot). Next, we tune $\xi_0$ to capture the spread in CDW melting temperatures in TiSe$_2$ flakes and find that $\xi_0$ falls within 10 nm and 50 nm. If we consider only the data points with circularity near 1 (blue data points in Fig. 3), we find that the correlation length lies between 10 and 20 nm. Thus, $\xi_0$ is on the same order of magnitude to the reported CDW correlation length in 1$T$-TiSe$_2$, which was approximated by the CDW domain size.[19]

Just like any phase transitions, the domain size of CDW is governed by the nucleation process of CDW. In TiSe$_2$, STM studies have shown the presence of agglomerated Ti intercalants that act as nucleation sites for CDW formation.[16] The Ti intercalants induce displacements in the Se atomic positions, which favor the CDW formation. The CDW nuclei grow into CDW domains, with the domain boundary having phase slips.

We thus look for the presence of Ti self-intercalants in the <100> zone axis of bulk 1$T$-TiSe$_2$. The atomic-resolution HAADF-STEM image (Fig. 4a) shows that 1$T$-TiSe$_2$ contains distorted regions where the atomic columns appear blurred. Intensity line profiles across the vdW gap show a small intensity peak in the vdW gap in the distorted region, which is absent in the undistorted region

(Fig. 4b). This peak indicates the presence of an intercalant in the vdW gap. Several intercalants are present in the distorted regions, as indicated by white dotted circles in Fig. 4a. These intercalants introduce a strain field that distorts the atomic positions, which leads to the blurring of atomic columns in the HAADF-STEM image. Hence, we confirm the presence of intercalants in TiSe$_2$, which act as nucleation sites for CDW formation from STM studies.[16]

Next, we measure the distance between the distorted regions of the 1T-TiSe$_2$ by analyzing a larger field-of-view, to assess whether the length scale between nucleating sites matches the domain size and $\xi_0$. We focus on the intercalant distance in the same layer to isolate the in-plane component of the correlation length from the out-of-plane component, as we are interested in the lateral confinement effects on the CDW formation. Figure 4c shows two such distances of 58 nm and 28 nm between intercalated regions, and Supplementary Figure 9 presents a 32 nm spacing from a different region of the bulk sample. These measurements confirm that the separation between the intercalated regions is on the order of tens of nanometers, similar to $\xi_o$ of 10–50 nm obtained from our G-L calculations. We thus conclude this represents the mean distance between nucleation sites.

As the nanoflake dimensions approach $\xi_0$, the G-L theory predicts CDW will be unstable with the transition temperature essentially approaching zero. Figure 3 shows this with the calculated CDW transition temperature suppressed to 0 K for systems with $R < \xi_o$. Indeed, we observed that some nanoflakes with small surface areas did not transition into a CDW phase at the base temperature of ~ 100 K, using a liquid nitrogen TEM holder. Even with a liquid helium TEM holder, two of the tested flakes did not show CDW ordering at the base temperature of 20 K (represented with stars in Figure 3), which suggests a complete suppression of CDW formation for these two TiSe$_2$ nanoflakes.

Microscopically, the complete suppression of CDW in smaller TiSe$_2$ nanoflakes is attributed to defect starvation. As the nanoflake size decreases, the number of Ti intercalant clusters also decreases. For flake sizes below the mean cluster spacing, some nanoflakes will not have any clusters, i.e., CDW nucleation sites, thereby preventing CDW formation. Inter-cluster distances of ~ 50 nm (Fig. 4c) agrees with the R of < 50 nm from the G-L theory.

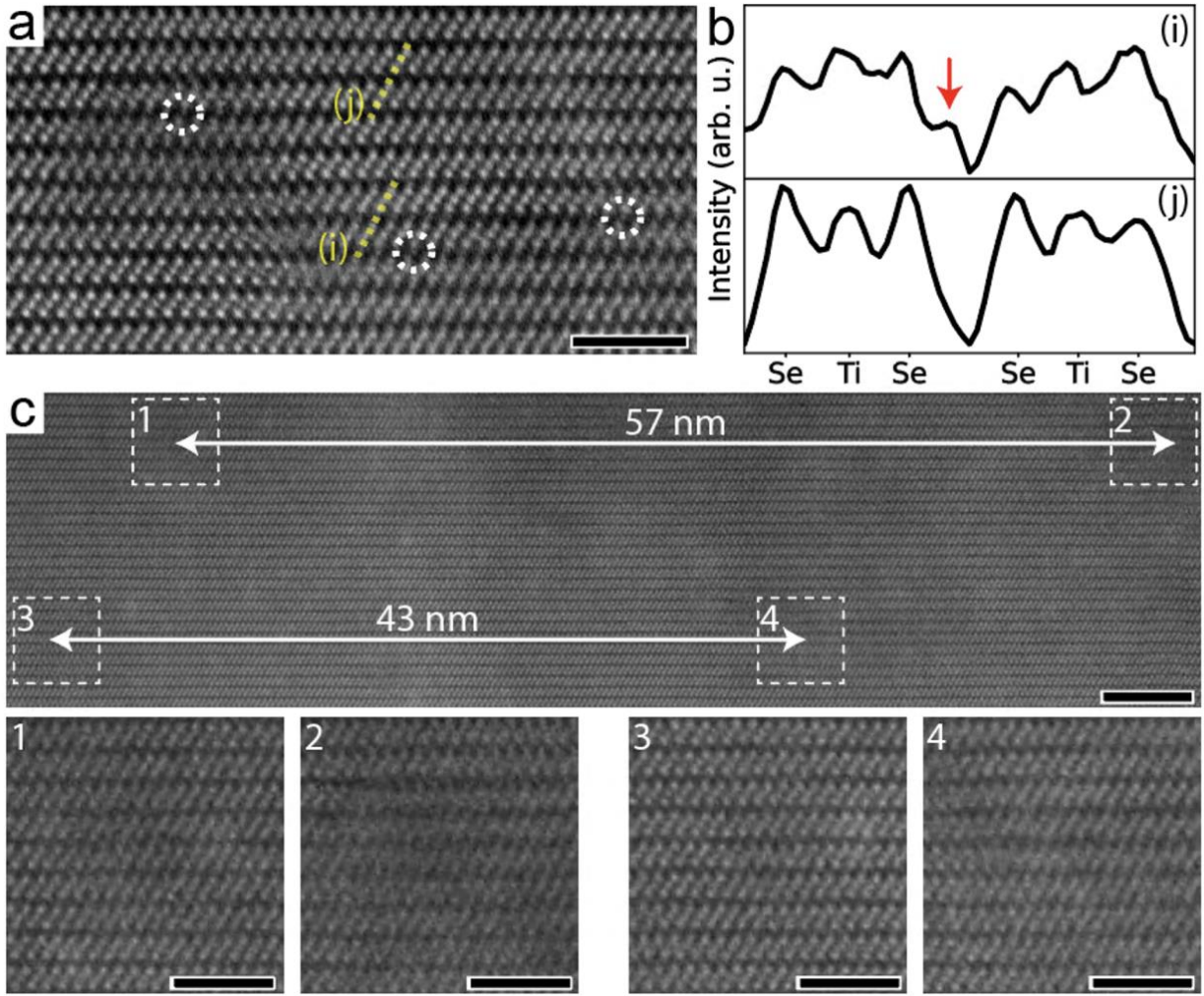

**Figure 4: Ti self-intercalants in TiSe. a.** HAADF-STEM image of 1*T*-TiSe$_2$ in the [100] projection. White dotted circles show intensity in the vdW gap from the intercalants. The atomic columns get distorted and appear blurred in the presence of intercalants. **b.** Intensity line profiles of atomic columns indicated by *(i)* and *(j)* in panel **(a)**. Line profile *(i)* shows a peak (red arrow) in the vdW gap indicating an intercalant. In contrast, line profile *(j)*, taken from a region without intercalant, does not show a peak in the vdW gap. **c.** A low-magnification atomic-resolution image showing the distance between distorted regions, i.e., CDW nucleation sites. Panels **1–4** show zoomed-in view of the distorted regions. Scale bars for **(a)** and panels **1–4** = 2 nm, for **(c)** = 5 nm.

Overall, among all the flakes examined, we successfully observed the melting of the CDW phase in 19 nanoflakes. Although these nanoflakes vary in thickness, the CDW transition temperature in TiSe$_2$ remains independent of thickness, staying between 210–230 K as long as the flakes are laterally large.[17] Thus, variations in thickness do not significantly influence $T_{melt}$ values plotted in Figure 3.

Finally, we note that all the temperature readings reported in this work are taken from the sensors on TEM holders, and do not necessarily represent the actual sample temperature. Testing on larger flakes (effective radii > 200 nm), which should have CDW melting temperatures comparable to bulk, shows that the temperature controller for the liquid nitrogen holder is accurate to within 10–20 K. For the liquid helium holder, a post-experiment measurement with a temperature sensor near the sample shows that the sample temperature was ~ 15 K higher than the controller's reported temperature. The noted temperature difference between the sensor and samples does not change the central observation of melting point depression of the CDW due to lateral confinements.

**Conclusion**

In conclusion, we find that the CDW melting temperature decreases and can be completely suppressed as the lateral size of 1$T$-TiSe$_2$ flakes decreases. The observed melting point depression and absence of CDWs is well modeled by the G-L theory. Microscopically, we relate the correlation length obtained from the CDW melting behavior to the distance between defects (Ti intercalants) that occur during the growth process. Our findings indicate that careful defect engineering can be an effective tool to enhance the stability and nucleation of strongly correlated states at the nanoscale. Technologically, our work suggests a minimum device size that can exploit CDW properties, which has implications for scalability in microelectronic applications.

**Acknowledgements**

The cryo-STEM experiments, theoretical modeling, and O.L. and E.-A.K. were supported by the Department of Energy, Basic Energy Sciences program Grant No. DE-SC0023905. O.L. is also supported by a Bethe-KIC postdoctoral fellowship at Cornell University. E.-A.K. was also funded in part by Gordon and Betty Moore Foundation's EPiQS Initiative, Grant GBMF10436 to E.-A.K. This research was funded in part by a QuantEmX grant from ICAM and the Gordon and Betty Moore Foundation through Grant GBMF9616, which supported travel to the University of Michigan for S.S. The Gordon & Betty Moore EPiQS Initiative, grant GBMF9062.01, supported synthesis of TiSe$_2$ bulk. This work made use of electron microscopes supported by the Platform for the Accelerated Realization, Analysis, and Discovery of Interface Materials (PARADIM) Award No. DMR-2039380, and the Cornell Center for Materials Research shared instrumentation facility. The instrument acquisitions were supported by the NSF (DMR-2039380). R.H. acknowledges support from the U.S. Department of Energy, Basic Energy Sciences, under award DE-SC0024147. M.G. recognizes support from the National Science Foundation (NSF) Small Business Innovation Research (SBIR) under award 2322155.